\documentclass[a4paper,11pt]{article}
\usepackage{pos1}

\usepackage{graphicx,times,amsmath,amsfonts,bbm,stmaryrd,simplewick,mathrsfs,dsfont,slashed}
\usepackage{hyperref}

\newcommand{\C}{\mathds C}
\newcommand{\N}{\mathds N}
\newcommand{\RR}{\mathds R}

\newcommand{\A}{\widetilde A}
\newcommand{\F}{\widetilde F}
\newcommand{\R}{\widetilde R}
\newcommand{\cG}{\mathcal{G}}

\def\ve{\varepsilon}
\def\im{\mathrm{i}}
\def\ep{\mathrm{e}}
\def\pa{\partial}
\def\diff{\mathrm{d}}
\def\md{\mathrm{d}}
\def\tr{\mathrm{tr}}
\def\sfrac#1#2{{\textstyle\frac{#1}{#2}}}
\def\>{\rangle}
\def\<{\langle}
\def\+{\dagger}

\def\={\ =\ }
\def\unity{\mathbbm{1}}
\def\und{\quad\textrm{and}\quad}
\def\with{\quad\textrm{with}\quad}

\title{The Nicolai map for super Yang--Mills theory \\
and application to the supermembrane}

\author*[a]{Olaf Lechtenfeld}

\affiliation[a]{Institut f\"ur Theoretische Physik, Leibniz Universit\"at Hannover,\\
  Appelstrasse 2, Hannover, Germany}

\emailAdd{olaf.lechtenfeld@itp.uni-hannover.de}

\abstract{
The nonlocal bosonic theory obtained from integrating out all anticommuting and auxiliary variables 
in a globally supersymmetric theory is characterized by the Nicolai map. We present a universal formula 
for the latter in terms of an ordered exponential of the integrated coupling flow operator, 
which can be canonically constructed. Also for supersymmetric gauge theories, this allows us to 
perturbatively construct the Nicolai map explicitly in terms of tree diagrams. For off-shell supersymmetry 
this works in any gauge, in the on-shell case the Landau gauge is required. The dimensional reduction 
of $D{=}10$ super Yang--Mills to maximally supersymmetric SU($N$) matrix mechanics (the BFSS model)
is known to provide a regularization of the $D{=}11$ supermembrane in light-cone gauge via its
incarnation as a one-dimensional gauge theory of area-preserving membrane diffeomorphisms.
We show how a well-defined corresponding Nicolai map perturbatively linearizes the supermembrane
in the small-tension regime and points the way for a computation of vertex-operator correlators.}

\FullConference{%
  Corfu Summer Institute 2021 "School and Workshops on Elementary Particle Physics and Gravity"\\
  29 August - 9 October 2021\\
  Corfu, Greece
}

\begin{document}
\maketitle

\section{Definition and construction of the Nicolai map}
\noindent
The key idea is best illustrated by an example. 
Let us look at the Wess--Zumino model in $3{+}1$ dimensional Minkowski space, 
consisting of a complex scalar~$\phi$, a Weyl fermion~$\psi$ and a complex auxiliary~$F$, 
characterized by a superpotential~$W(\phi)$ and featured in the off-shell lagrangian~\footnote{
A multi-field generalization is straightforward.}
\begin{equation}
{\cal L} \= \pa_\mu\phi^*\pa^\mu\phi + F^*F + 
\tfrac{\im}{2} \bar\psi\bar\sigma{\cdot}\pa\psi - \tfrac{\im}{2} \psi\sigma{\cdot}\pa\bar\psi +
W'(\phi)\,F + W'(\phi)^*F^* - \tfrac12\psi\,W''(\phi)\,\psi - \tfrac12 \bar\psi\,W''(\phi)^*\bar\psi\ ,
\end{equation}
where $\sigma=(\unity,\vec{\sigma})$ and $\bar\sigma=(\unity,-\vec{\sigma})$ 
with Pauli matrices~$\vec\sigma$.
Integrating out the auxiliary fields yields $F^*=-W'(\phi)$ and
\begin{equation}
{\cal L}_{\textrm{SUSY}} \= \bigl|\pa\phi\bigr|^2 -\bigl|W'(\phi)\bigr|^2 + 
\bigl( \tfrac{\im}{2} \bar\psi\,\bar\sigma{\cdot}\pa\psi - \tfrac12\psi\,W''(\phi)\,\psi + \textrm{h.c.}\bigr)\ .
\end{equation}
Integrating out the fermions $(\psi,\bar\psi)$ produces a functional determinant
$\det M=\exp\{\tfrac{\im}{\hbar}{\cdot}(-\im\hbar\,\tr\ln M)\}$
so that the action becomes
\begin{equation}
S_g[\phi] \= \smallint\!\diff^4\!x\ \bigl\{ |\pa\phi|^2 - |W'|^2 \bigr\}
\ -\ \im\hbar\,\tr\ln \Bigl(\begin{smallmatrix} 
W'' & \im\,\sigma{\cdot}\pa \\[4pt]
-\im\,\bar\sigma{\cdot}\pa & {W''}^* 
\end{smallmatrix} \Bigr)
\ =:\ S_g^b[\phi] + \hbar\,S_g^f[\phi]\ .
\end{equation}
Here, $g$ denotes some coupling constant(s) or parameter(s) inside the superpotential~$W(\phi)$.
The objects of desire are quantum correlators
\begin{equation} \label{Ycorr}
\bigl\< Y[\phi] \bigr\>_g \= \int\!\!{\cal D}\phi\ \ep^{\frac{\im}{\hbar} S_g[\phi]}\ Y[\phi]
\quad\with\quad \bigl\<\unity\bigr\> \= 1
\end{equation}
for any bosonic (local or nonlocal) functional~$Y$.

The path integral in (\ref{Ycorr}) describes a purely bosonic nonlocal field theory.
What is characteristic of its supersymmetric origin? In other words:
given such a nonlocal action~$S_g$, how could one infer its hidden supersymmetric root?
This question was answered in 1980 by Hermann Nicolai~\cite{Nic1,Nic2,Nic3}:
Such hiddenly supersymmetric theories admit a nonlocal and nonlinear invertible map
\begin{equation} \label{Nicdef}
T_g:\ \phi\,\mapsto\ \phi'[\phi;g]
\qquad\textrm{such that}\qquad
\bigl\<Y[\phi]\bigr\>_g \= \bigl\<Y[T_g^{-1}\phi]\bigr\>_0 \quad \forall\,Y\ ,
\end{equation}
relating correlators in the interacting theory ($g{\neq}0$) to (more complicated) correlators
in the free theory ($g{=}0$). For the path integrals, this is equivalent to
\begin{equation}
{\cal D}\phi\ \exp\bigl\{ \tfrac{\im}{\hbar} S_g[\phi] \bigr\} \=
{\cal D}(T_g\phi)\ \exp\bigl\{ \tfrac{\im}{\hbar} S_0[T_g\phi] \bigr\} \=
{\cal D}\phi\ \exp\bigl\{ \tfrac{\im}{\hbar} S_0[T_g\phi] + \tr\ln\tfrac{\delta T_g\phi}{\delta\phi} \bigr\}\ .
\end{equation}
Separating powers of $\hbar$ in the exponent, this splits into two properties,
\begin{subequations}
\begin{equation} \label{freeaction}
S_0^b[T_g\phi] \= S_g^b[\phi] 
\qquad\textrm{``free action condition''} \ ,
\end{equation}
\begin{equation} \label{detmatching}
{}\quad S_0^f - \im\,\tr\ln\tfrac{\delta T_g\phi}{\delta\phi} \= S_g^f[\phi]
\qquad\textrm{``determinant matching condition''} \ .
\end{equation}
\end{subequations}
Every Nicolai map has to fulfil these two conditions, which originally were taken as its definition.
The reason for the name of~(\ref{detmatching}) is that its exponentiation gives an equality 
of the functional fermion determinant~$\det M$ with the Jacobian of the transformation
(the first term is a constant since $S_0^f$ does not depend on~$\phi$).
From now on we put $\hbar{=}1$.

In 1984, the author derived (for his dissertation) an infinitesimal version~\cite{L1,DL1,DL2,L2} 
of the Nicolai map by considering the
$g$-derivative of~(\ref{Nicdef}),
\begin{equation} \label{Nicinf}
\begin{aligned}
\pa_g\bigl\<Y[\phi]\bigr\>_g 
&\ \stackrel{(\ref{Nicdef})}{=}\ \pa_g \bigl\<Y[T_g^{-1}\phi]\bigr\>_0 \\
&\=\, \bigl\<\pa_g Y[\phi]\bigr\>_g\ +\ 
\bigl\< \smallint (\pa_g T_g^{-1}\phi)\cdot\tfrac{\delta Y}{\delta\phi}[T_g^{-1}\phi]\bigr\>_0 \\
&\!\stackrel{(\ref{Nicdef})^{-1}}{=} \bigl\<\pa_g Y[\phi]\bigr\>_g\ +\ 
\bigl\< \smallint (\pa_g T_g^{-1}\circ T_g)\phi\cdot\tfrac{\delta Y}{\delta\phi}[\phi]\bigr\>_g
\ =:\ \bigl\< \bigl(\pa_g + R_g[\phi]\bigr)\,Y[\phi] \bigr\>_g
\end{aligned}
\end{equation}
with a ``flow operator''~\footnote{
We write $\diff x$ for the spacetime volume differential as long as its dimension remains unspecified.}
\begin{equation}
R_g[\phi] \= \int\!\diff x\ \bigl( \pa_g T_g^{-1}\circ T_g\bigr)\phi(x)\,\frac{\delta}{\delta\phi(x)}
\end{equation}
representing a functional differential operator derived from~$T_g$.

Nothing is gained, however, by these formal considerations, unless we can reverse the logic
and somehow obtain~$R_g$ and exponentiate it in order to create a finite flow~$T_g$ from
$g'{=}0$ to $g'{=}g$, by inverting
\begin{equation}
\bigl(T_g^{-1}\phi\bigr)(x) \= \exp\bigl\{ g\,\bigl(\pa_{g'}+R_{g'}[\phi]\bigr)\bigr\}\,\phi(x)\,\big|_{g'=0}
\= \smallsum_{n=0}^\infty \tfrac{g^n}{n!}\,\bigl( \pa_{g'} + R_{g'}[\phi] \bigr)^n\ \phi(x)\,\big|_{g'=0} \ .
\end{equation}
At this stage two remarks are in order.
Firstly, $R_g$ is a derivation, and hence $T_g^{-1}$ acts distributively,
\begin{equation}
R_g\,Y[\phi] \= \smallint \tfrac{\delta Y}{\delta\phi}\cdot R_g\phi 
\qquad\Leftrightarrow\qquad
T_g^{-1} Y[\phi] \= Y[T_g^{-1}\phi] \ .
\end{equation}
Secondly, by moving the map ``to the other side'',
\begin{equation}
\bigl\<Y[\phi]\bigr\>_0 \= \bigl\<Y[T_g\phi]\bigr\>_g\ ,
\end{equation}
choosing $\pa_g Y=0$ and differentiating with respect to~$g$, we learn that
\begin{equation}
0 \= \pa_g \bigl\<Y[T_g\phi]\bigr\>_g 
\ \stackrel{(\ref{Nicinf})}{=}\ \bigl\<\bigl(\pa_g+R_g[\phi]\bigr)\,Y[T_g\phi] \bigr\>_g
\= \bigl\< \smallint \bigl(\pa_g+R_g[\phi]\bigr)\,T_g\phi \cdot \tfrac{\delta Y}{\delta\phi} [T_g\phi]\bigr\>_g
\end{equation}
for any (not explicitly $g$-dependent) functional~$Y$, and therefore
\begin{equation} \label{fixpoint}
\bigl(\pa_g + R_g[\phi]\bigr)\,T_g\phi(x) \= 0\ .
\end{equation}
This ``fixpoint property'' of the Nicolai map under the infinitesimal flow allows us
to directly construct $T_g\phi$ from $R_g$ without invoking the inverse first.

Indeed, (\ref{fixpoint}) is formally solved by a path-ordered exponential,
\begin{equation} \label{universal}
T_g\phi \= {\cal P} \exp \Bigl\{-\!\!\int_0^g\!\!\diff h\ R_h[\phi]\Bigr\}\ \phi
\= \sum_{s=0}^\infty (-1)^s\!\!\int_0^g\!\!\diff h_s \ldots \!\int_0^{h_3}\!\!\!\!\diff h_2 \!\int_0^{h_2}\!\!\!\!\diff h_1\
R_{h_s}[\phi] \ldots R_{h_2}[\phi]\,R_{h_1}[\phi]\ \phi\ ,
\end{equation}
providing a ``universal formula'' for the Nicolai map in terms of the infinitesimal coupling flow~\cite{LR1}.
It is often useful to expand the flow operator in powers of the coupling,
\begin{equation}
R_g[\phi] \= \sum_{k=1}^\infty g^{k-1} r_k[\phi] \= r_1[\phi] + g\,r_2[\phi] + g^2 r_3[\phi] + \ldots
\end{equation}
from which one easily computes a power series expansion for the map itself,
\begin{equation}
T_g\phi \= \!\sum_{\bf n} g^n\,c_{\bf n}\,r_{n_s}[\phi]\ldots r_{n_2}[\phi]\,r_{n_1}[\phi]\ \phi
\qquad\textrm{with}\quad
{\bf n} = (n_1,n_2,\ldots,n_s)\ ,\quad n_i\in\N \ ,\quad \smallsum_i n_i = n\ ,
\end{equation}
where $1\le s \le n$ and the $n{=}0$ term is the identity.
The numerical coefficients are computed as
\begin{equation}
c_{\bf n} \ = (-1)^s\!\!\int_0^1\!\!\diff x_s\;x_s^{n_s-1} \ldots 
\!\!\int_0^{x_3}\!\!\!\!\diff x_2\;x_2^{n_2-1} \!\!\int_0^{x_2}\!\!\!\!\diff x_1\;x_1^{n_1-1} \=
(-1)^s\bigl[ n_1\cdot(n_1+n_2)\cdots(n_1+n_2+\ldots+n_s)\bigr]^{-1}
\end{equation}
and related to the Stirling numbers of the second kind.
Writing out the first few terms, the perturbative Nicolai map reads
\begin{equation}
\begin{aligned}
T_g\phi &\= \phi \ -\ g\,r_1 \phi \ -\ \sfrac12g^2\bigl(r_2-r_1^2\bigr)\phi\ -\ 
\sfrac16g^3\bigl(2r_3-r_1r_2-2r_2r_1+r_1^3\bigr)\phi \\
&\quad -\sfrac{1}{24}g^4\bigl(6r_4-2r_1r_3-3r_2r_2+r_1^2r_2-6r_3r_1
+2r_1r_2r_1+3r_2r_1^2-r_1^4\bigr)\phi \ +\ {\cal O}(g^5)\, .
\end{aligned}
\end{equation}

For computing correlation functions \`a la (\ref{Nicdef}) we need the inverse map.
It possesses an analogous universal representation in terms of an anti-path-ordered exponential,
which gives rise to a different power series expansion,
\begin{equation}
T_g^{-1}\phi \= \!\sum_{\bf n} g^n\,d_{\bf n}\,r_{n_s}[\phi]\ldots r_{n_2}[\phi]\,r_{n_1}[\phi]\ \phi
\quad\textrm{with}\quad
c_{\bf n}  \= \bigl[ n_s\cdot(n_s+n_{s-1})\cdots(n_s+n_{s-1}+\ldots+n_1)\bigr]^{-1}
\end{equation}
whose first terms are
\begin{equation}
\begin{aligned}
T_g^{-1}\phi &\= \phi \ +\ g\,r_1 \phi \ +\ \sfrac12g^2\bigl(r_2+r_1^2\bigr)\phi\ +\ 
\sfrac16g^3\bigl(2r_3+2r_1r_2+r_2r_1+r_1^3\bigr)\phi \\
&\quad +\sfrac{1}{24}g^4\bigl(6r_4+6r_1r_3+3r_2r_2+3r_1^2r_2+2r_3r_1
+2r_1r_2r_1+r_2r_1^2+r_1^4\bigr)\phi \ +\ {\cal O}(g^5)\, .
\end{aligned}
\end{equation}

Still, we have to establish the existence of the flow operator~$R_g$ 
and find an explicit expression for it. We shall do this now for the exemplary case
of scalar theories (gauge theories will be treated in the following section).
If supersymmetry is realized off-shell on the action~$S$ then there exists a functional
$\mathring{\Delta}_\alpha[\phi,\psi,F]$
such that
\begin{equation}
\pa_g S[\phi,\psi,F] \= \delta_\alpha \mathring{\Delta}_\alpha[\phi,\psi,F]
\end{equation}
for the supersymmetry transformations~$\delta_\alpha$, 
where $\alpha$ denotes a Majorana spinor index.
Integrating out the auxiliary~$F$ one has that
\begin{equation} \label{dgS}
\pa_g S_{\textrm{SUSY}}[\phi,\psi] \= \delta_\alpha \Delta_\alpha[\phi,\psi]
\qquad\textrm{with}\quad 
\Delta_\alpha[\phi,\psi] = \mathring{\Delta}_\alpha[\phi,\psi,-\smash{W'}^*(\phi)]
\end{equation}
for the on-shell action $S_{\textrm{SUSY}}\,{=}\int\!\!\diff x\,{\cal L}_{\textrm{SUSY}}$
with an anticommuting functional~$\Delta_\alpha$.
For our Wess--Zumino model example, it reads
$\Delta_\alpha=\tfrac12\int\!\diff^4\!x\,\psi_\alpha\,\pa_gW'(\phi)$.
The construction of~$R_g$ employs the supersymmetry Ward identity:
\begin{equation}
\pa_g \!\int\!\!{\cal D}\phi \!\int\!\!{\cal D}\psi\ \ep^{\im S_{\textrm{\tiny SUSY}}[\phi,\psi] }\ Y[\phi]
\= \int\!\!{\cal D}\phi \!\int\!\!{\cal D}\psi\ \ep^{\im S_{\textrm{\tiny SUSY}}[\phi,\psi] }\
\bigl( \pa_g + \im\Delta_\alpha[\phi,\psi]\ \delta_\alpha\bigr) Y[\phi]\ ,
\end{equation}
Integrating out the fermions contracts bilinears to produce fermion propagators
$\bcontraction{}{\psi}{\,}{\psi} \psi\,\psi$ (in the $\phi$ background), hence~\cite{L1}
\begin{equation}
R_g[\phi] \= \im\,\bcontraction{}{\Delta}{_\alpha[\phi]\ }{\delta} \Delta_\alpha[\phi]\ \delta_\alpha \=
\im\int\!\diff x\ \bcontraction{}{\Delta}{_\alpha[\phi]\ }{\delta} \Delta_\alpha[\phi]\ \delta_\alpha \phi(x)\ 
\frac{\delta}{\delta\phi(x)}
\end{equation}

For a simple example of the Wess--Zumino model with (massless) superpotential 
$W=\tfrac13 g\phi^3$, one finds that
\begin{equation}
R_g[\phi] \= \tfrac{\im}{2} \smallint\!\!\smallint\diff^4\!x\,\diff^4\!y\ 
\bigl\{ \phi^2(x)\,\bcontraction{}{\psi}{(x)\ }{\psi} \psi(x)\ \psi(y) 
\ +\ {\smash{\phi^*}}^2(x)\,\bcontraction{}{\bar\psi}{(x)\ }{\psi} \bar\psi(x)\ \psi(y) \bigr\}_{\alpha\alpha}\,
\tfrac{\delta}{\delta\phi(y)} \ -\ \textrm{h.c.}
\end{equation}
where the subscript on the curly brace indicates a spin trace.
It is instructive to develop a diagrammatical shorthand notation.
For the sake of illustration, here we oversimplify $(\phi,\phi^*)\sim\phi$ and write \\
\centerline{\includegraphics[width=0.7\paperwidth]{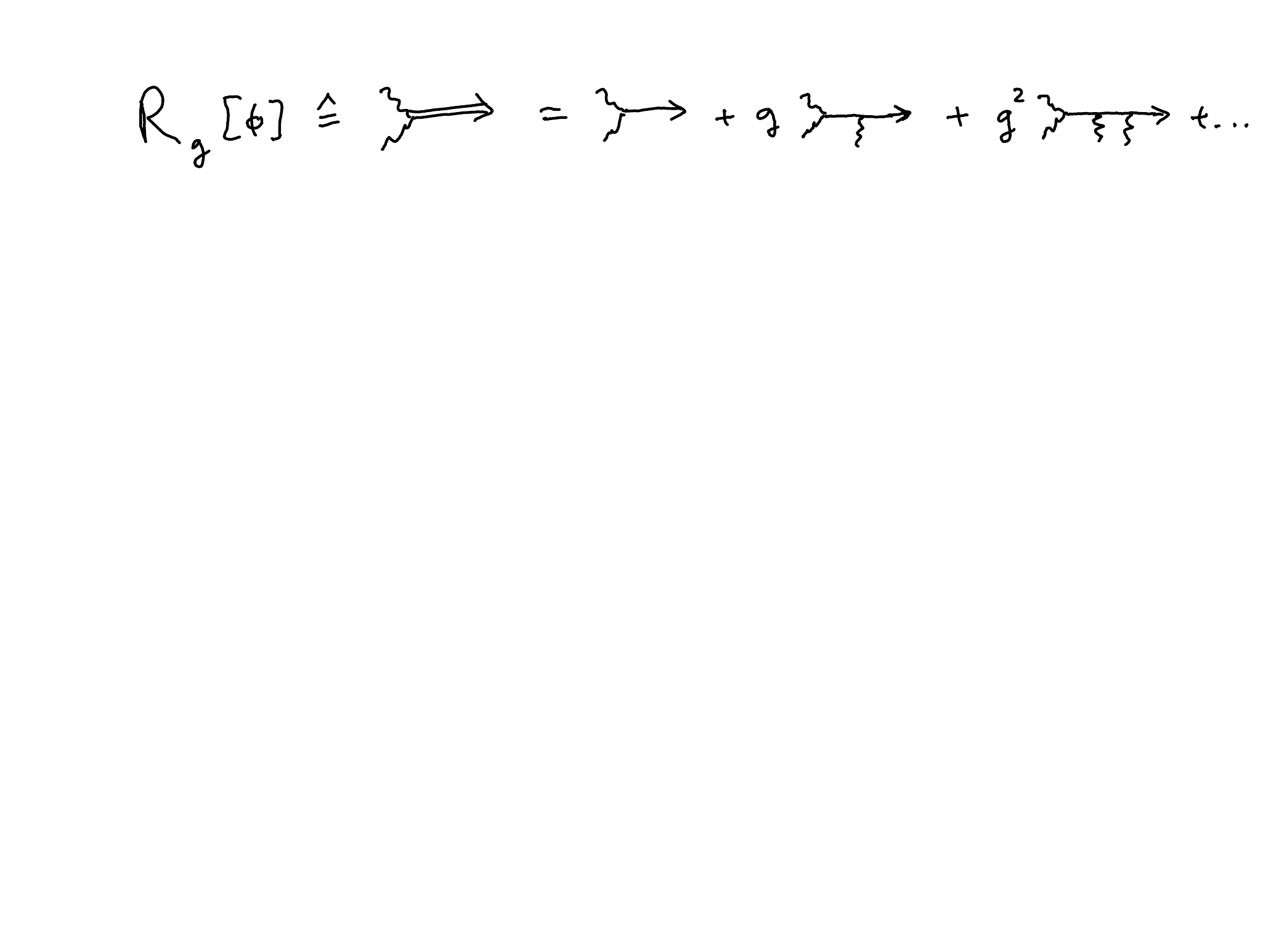}} \\
with the graphical rules~\cite{FL} \\[2pt]
\centerline{\includegraphics[width=0.7\paperwidth]{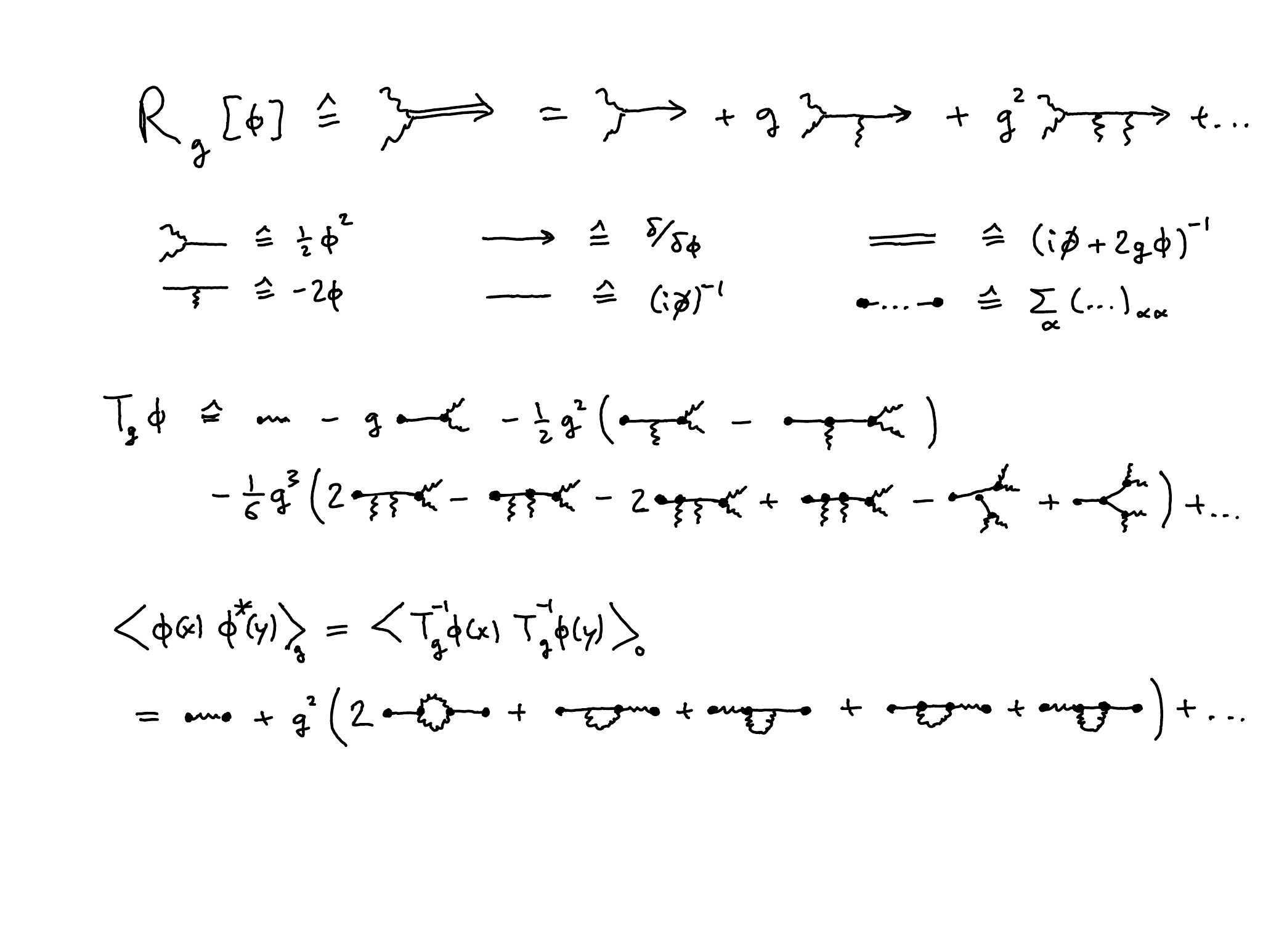}} \\
The linear tree for $R_g$ exponentiates to a series of branched trees for $T_g\phi$, \\[2pt]
\centerline{\includegraphics[width=0.7\paperwidth]{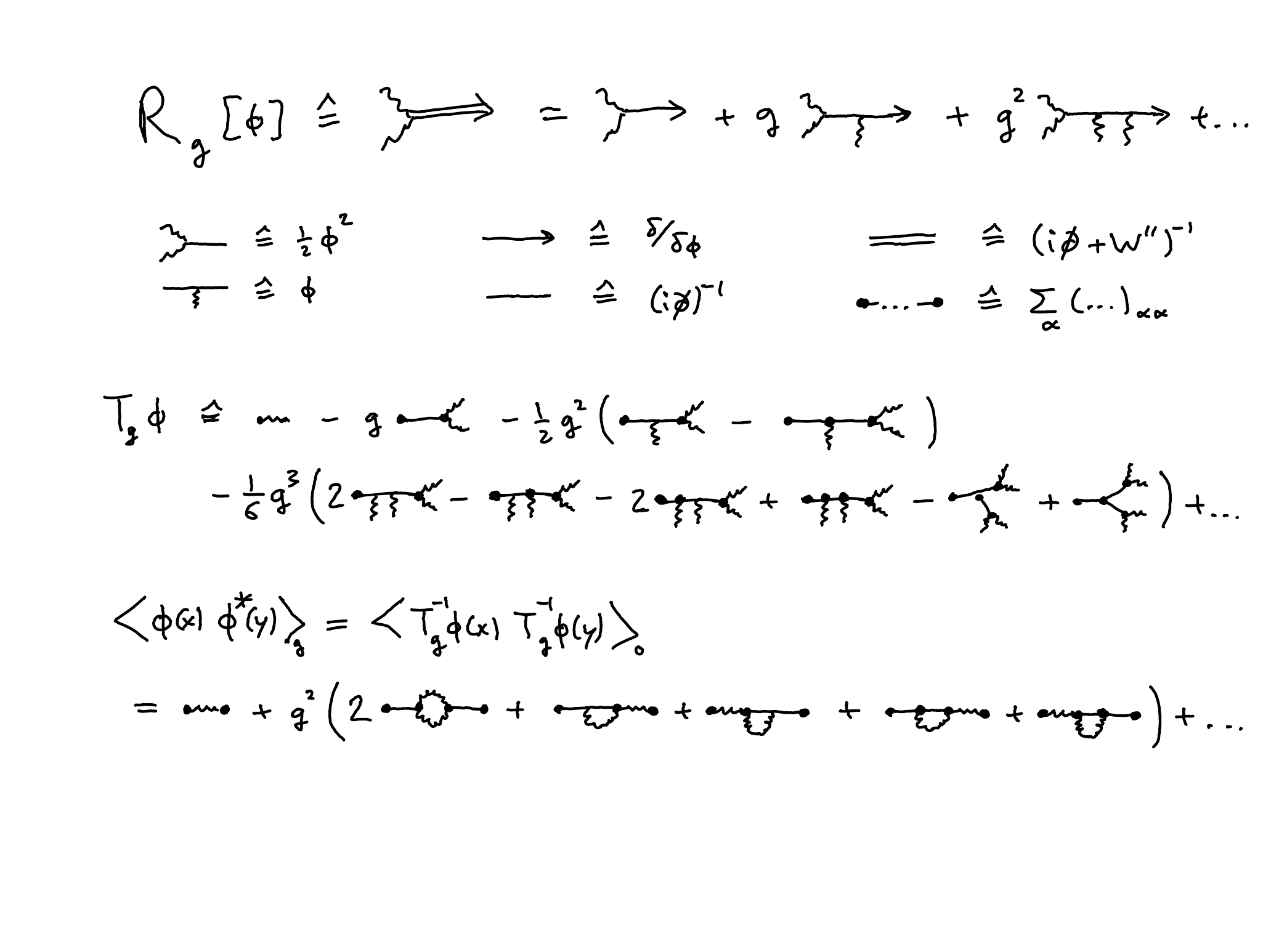}} \\
and likewise for the inverse~$T_g^{-1}\phi$. 
Inserting the latter into (\ref{Nicdef}) and performing the free-theory bosonic contractions,
one obtains an alternative Feynman perturbation series for correlators, as displayed here
for the two-point function: \\[2pt]
\centerline{\includegraphics[width=0.7\paperwidth]{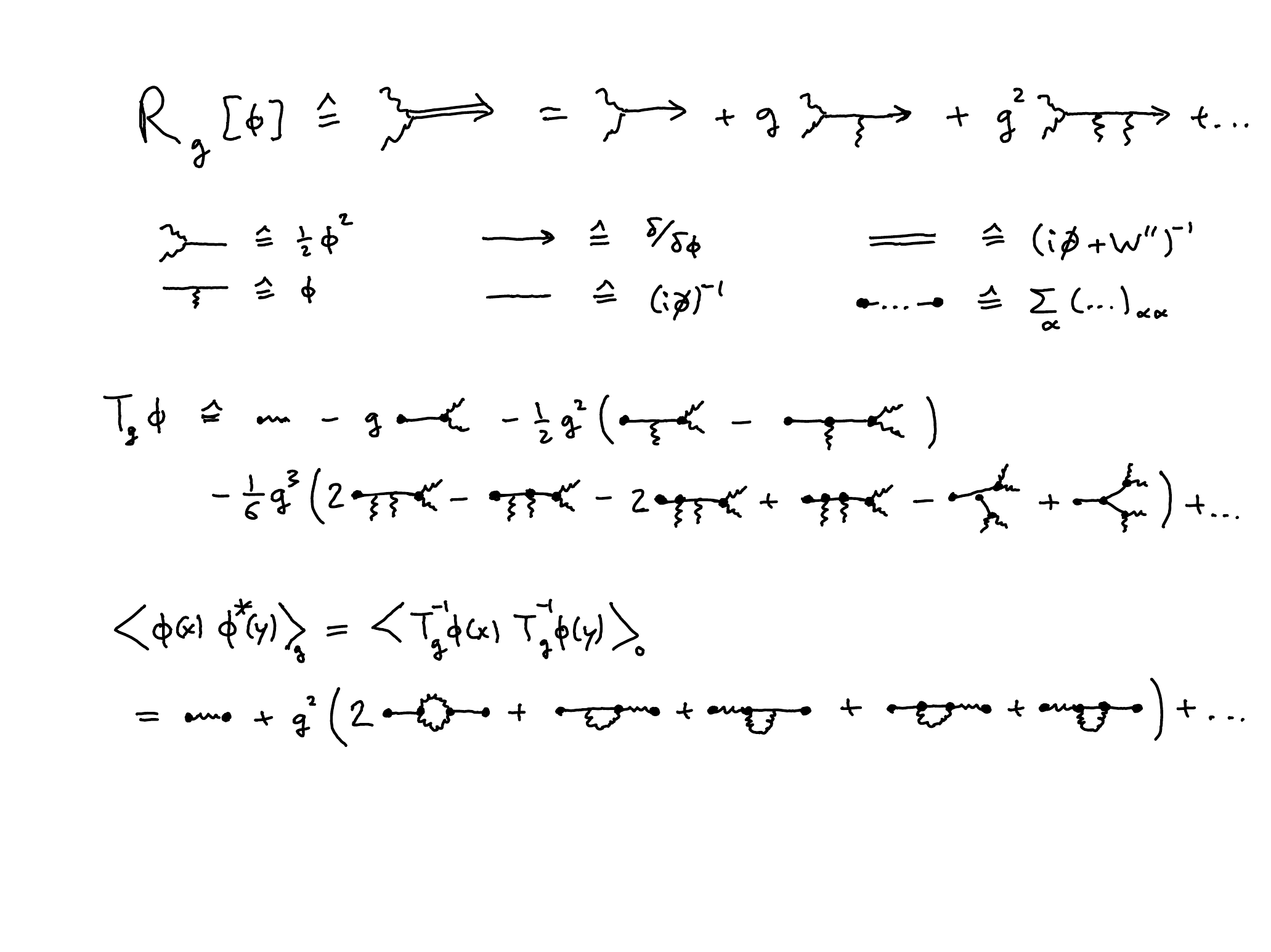}} \\
Notably, the multiple action of~$R_g$ produces multiple spin traces
(graphically separated by dots).
The supersymmetric cancellation of the leading UV divergencies is automatially built in,
as pure fermion loops are absent as well as boson tadpoles.

\section{The case of gauge theories}
\noindent
Suprsymmetric gauge theories present additional challenges. 
Firstly, one has to deal with the gauge redundancy
necessitating a (supersymmetry-breaking) gauge fixing and, secondly, the $g$-derivative of the 
supersymmetric action cannot easily be expressed as a supervariation. 
We eliminate the auxiliary field ($D$-term), 
use a local gauge-fixing functional $\cG$ to fix a gauge $\cG(A){=}0$ with a parameter~$\xi$
and include the corresponding ghost fields to formulate a BRST-invariant on-shell action
\begin{equation}
S_{\textrm{SUSY}}[A,\lambda,c,\bar c] \= \int\!\diff x\ \tr\bigl\{ 
-\tfrac14 F_{\mu\nu}F^{\mu\nu} -\tfrac{1}{2\xi}\cG(A)^2
-\tfrac{\im}{2}\bar\lambda\slashed{D}\lambda 
+ \bar{c}\,\tfrac{\pa\cG}{\pa A_\mu}D_\mu c \bigr\}
\end{equation}
for $su(N)$-valued gluons~$A_\mu{=}A_\mu^A T^A$, 
gluinos~$\lambda_\alpha{=}\lambda_\alpha^A T^A$,
a ghost $c{=}c^A T^A$ and an antighost $\bar c{=}\bar c^A T^A$,
with $F_{\mu\nu}{=}\pa_\mu A_\nu{-}\pa_\nu A_\mu{+}g[A_\mu,A_\nu]{=}F_{\mu\nu}^AT^A$ 
and group generators subject to $[T^A,T^B]{=}f^{ABC}T^C$ with $A,B,\ldots=1,2,\ldots,N^2{-}1$.
The trace refers to the color degrees of freedom.
We allow for various spacetime dimensionalities~$D$ by letting the fields live on~$\RR^{1,D-1}$
so that $\mu,\nu,\ldots=0,1,\ldots,D{-}1$ and $\alpha=1,\ldots,r$,
where $r$ is the complex dimension of the corresponding Majorana representation,
i.e.~$\lambda^A\in\C^r$. It essentially grows exponentially with~$D$.
In the following, we present two different attempts to emulate the successful scalar-field procedure.

In version A~\cite{ALMNPP}, from $F=\diff A+g\,A{\wedge}A$ we see that $g{=}0$ is the free theory.
A quick computation shows things now are more involved than in~(\ref{dgS}),
\begin{equation}
\pa_g S_{\textrm{SUSY}} \= \delta_\alpha \Delta_\alpha 
\ +\ q\smallint\tr\,\im\bar\lambda\slashed{A}\lambda
\ +\ \smallint\tr\,\bar{c}\tfrac{\pa\cG}{\pa A_\mu}A_\mu c
\qquad\textrm{with}\quad q=\tfrac{D-1}{r}-\tfrac12
\end{equation}
where ($\gamma^{\mu\nu}=\tfrac12[\gamma^\mu,\gamma^\nu]$)
\begin{equation}
\Delta_\alpha \= -\tfrac{1}{2r}\smallint\tr\,(\gamma^{\mu\nu}\lambda)_\alpha A_\mu A_\nu
\end{equation}
is the gauge-theory counterpart of the on-line functional in~(\ref{dgS}). However,
we have to fight with a ghost contribution and a ``mismatch''~$q$ in the construction of~$R_g$.
With the help of the broken-supersymmetry and BRST Ward identities one derives that
\begin{equation}
\pa_g\bigl\<Y[A]\bigr\>_g \= \bigl\< \bigl(\pa_g+R_g[A]+Z_g[A]\bigr)\,Y[A]\bigr\>_g
\end{equation} 
where
\begin{equation}
R_g \= \im\,\bcontraction{}{\Delta}{_\alpha\ }{\delta}\Delta_\alpha\ \delta_\alpha
\ -\ \bcontraction{}{\Delta}{_\alpha (}{\delta}\Delta_\alpha (\delta_\alpha
\bcontraction{}{\Delta}{_{\textrm{gh}})\,}{s}\Delta_{\textrm{gh}})\,s 
\qquad\textrm{with}\quad \Delta_{\textrm{gh}} = \smallint\tr\,\bar{c}\,\cG(A)\ ,
\end{equation}
\begin{equation}
Z_g \= (\bcontraction[1.5ex]{}{s}{\,\bcontraction{}{\Delta}{_\alpha)\,(}{\delta}\Delta_\alpha)\,(\delta_\alpha}{\Delta}
s\,\bcontraction{}{\Delta}{_\alpha)\,(}{\delta}\Delta_\alpha)\,(\delta_\alpha\Delta_{\textrm{gh}})
\ -\ q \smallint\tr\,\bcontraction{}{\bar\lambda}{\slashed{A}}{\lambda}\bar\lambda\slashed{A}\lambda
\ +\ \im\smallint\tr\,\bcontraction{}{\bar{c}}{\tfrac{\pa\cG}{\pa A_\mu}A_\mu}{c}
\bar{c}\tfrac{\pa\cG}{\pa A_\mu}A_\mu c \ , \qquad\quad{}
\end{equation}
and $s$ denotes the BRST (or Slavnov) variation.
The contractions signify gaugino or ghost propagators.
The multiplicative contribution~$Z_g$ destroys the derivation property of~$R_g$ and hence
the distributivity of~$T_g$, which is not acceptable.
A somewhat lengthy computation reveals, however, that in the Landau gauge,
$\cG{=}\pa^\mu A_\mu$ with $\xi{\to}\infty$, the obstacle may be overcome,
\begin{equation}
Z_g=0 \qquad\textrm{if and only if}\quad q=\tfrac1r \quad\Leftrightarrow\quad 
r=2(D{-}2) \quad\Leftrightarrow\quad  D=3,4,6,10\ .
\end{equation}
Amazingly, these are precisely the ``critial spacetime dimensions'' 
which admit super Yang--Mills theory to exist~\cite{BSS},
demonstrating that the Nicolai map knows about them~\cite{ANPP}!

For version B~\cite{L1,MN,LR2}, we restrict to a linear gauge 
$\cG(A)=n{\cdot}A$ or $\pa{\cdot}A$ and rescale all fields to tilded versions 
in order to pull out the gauge coupling. In particular,
\begin{equation}
g\,A=:\A \quad\Rightarrow\quad
S_{\textrm{SUSY}}[\A,\widetilde\lambda,\widetilde{c},\widetilde{\bar c}] 
\= \tfrac{1}{g^2}\!\!\int\!\!\diff x\ \tr\bigl\{ 
- \tfrac14 \F_{\mu\nu}\F^{\mu\nu} -\tfrac{1}{2\xi}\cG(\A)^2
- \tfrac{\im}{2}\widetilde{\bar\lambda}\widetilde{\slashed{D}}\widetilde{\lambda}
+ \sqrt{g}\,\widetilde{\bar{c}}\,\tfrac{\pa\cG}{\pa\A_\mu}\widetilde{D}_\mu \widetilde{c} \bigr\}
\end{equation}
where the tilded quantities are $g$-independent (or evaluated at $g{=}1$).
Since the $g$-derivative now is proportional to the action itself,\footnote{
except for the ghost term, which has to be scaled non-canonically}
we can use off-shell supersymmetry (only in $D{\le}4$ though) to obtain
\begin{equation}
\pa_g S_{\textrm{SUSY}} \= -\tfrac{1}{g^3} \,
\bigl\{ \delta_\alpha\widetilde\Delta_\alpha - \sqrt{g}\,\,s\,\widetilde\Delta_{\textrm{gh}} \bigr\}
\end{equation}
where
\begin{equation}
\widetilde\Delta_\alpha = -\tfrac{1}{2r}\smallint\tr\,(\gamma^{\mu\nu}\widetilde\lambda)_\alpha \F_{\mu\nu}
\quad\und\quad \widetilde\Delta_{\textrm{gh}} = \smallint\tr\,\widetilde{\bar{c}}\,\cG(\A)\ .
\end{equation}
Now we may proceed using broken-supersymmetry and BRST Ward identities to get
\begin{equation}
\pa_g \bigl\< Y[\A] \bigr\>_g \= \bigl\< \bigl(\pa_g + \R_g[\A] \bigr)\,Y[\A] \bigr\>_g
\end{equation}
where
\begin{equation}
\R_g \= -\im\,\bcontraction{}{\widetilde\Delta}{_\alpha\,}{\delta}\widetilde\Delta_\alpha\,\delta_\alpha
\ +\ \tfrac{\im}{\sqrt{g}}\,\bcontraction{}{\widetilde\Delta}{_{\textrm{gh}}\,}{s}\widetilde\Delta_{\textrm{gh}}\,s
\ -\ \tfrac{1}{\sqrt{g}}\,\bcontraction{}{\widetilde\Delta}{_\alpha (}{\delta}\widetilde\Delta_\alpha (\delta_\alpha
\bcontraction{}{\widetilde\Delta}{_{\textrm{gh}})\,}{s}\widetilde\Delta_{\textrm{gh}})\,s \ .
\end{equation}
Yet, in this version, we cannot expand around $g{=}0$ but for perutrbation theory must scale back to
\begin{equation}
A \= \tfrac1g\A \qquad\Rightarrow\qquad
R_g[A] \= \tfrac1g \bigl(\R_g[\A]-\smallint\A\tfrac{\delta}{\delta\A} \bigr)\ .
\end{equation}
Note that $\R_g[gA]\neq gR_g[A]$ but contains an Euler operator w.r.t.~$A$.
This is crucial to remove the formal $g{\to}0$ singularity in the above expression,
so that in fact $\lim_{g\to0}R_g$ is finite. 
We can give an explicit expression for any gauge but limited to $D{\le}4$:
\begin{equation}
\overleftarrow{R}_g[A] \= \tfrac{1}{2r} \smallint\!\!\smallint\!\!\smallint\tr\
\overleftarrow{\tfrac{\delta}{\delta A_\mu}}\,P_\mu^{\ \nu}\,\bigl\{
\gamma_\nu\,\bcontraction{}{\bar\lambda}{\ }{\lambda}
\bar\lambda\ \lambda\,\gamma^{\rho\sigma} A_\rho (A-2\pa\Box^{-1}\pa{\cdot}A)_\sigma \bigr\}_{\alpha\alpha}
+\ \smallint\!\!\smallint\tr\ \overleftarrow{\tfrac{\delta}{\delta A_\mu}}\,A_\mu\,\Box^{-1}\pa{\cdot}A
\ +\ O(\cG)
\end{equation}
with the non-Abelian transversal projector
\begin{equation}
P_\mu^{\ \nu} \= 
\delta_\mu^{\ \nu}\unity\ -\ D_\mu \bcontraction{}{c}{\ }{{\bar c}}c\ \bar c\,\tfrac{\pa\cG}{\pa A_\nu}
\qquad\Rightarrow\qquad
\tfrac{\pa\cG}{\pa A_\mu}\,P_\mu^{\ \nu} = 0 = P_\mu^{\ \nu} D_\nu
\end{equation}
forcing the flow onto the gauge surface: $R_g\cG\sim\cG$. 
For the Landau gauge, $\cG{=}\pa{\cdot}A$, all expressions simplify considerably.
We have reversed the direction of the derivatives since acting towards the left is more
convenient for the graphical representation.

So the upshot of both versions A and~B is that our explicit construction formula~(\ref{universal})
carries over to gauge theory, for $D{\le}4$ in any gauge and for $D{=}6$ and~$10$ in the Landau gauge,
\begin{equation} \label{universalA}
T_g A\= {\cal P} \exp \Bigl\{-\!\int_0^g\!\!\diff h\ R_h[A]\Bigr\}\ A
\= \sum_{\bf n} g^n\,c_{\bf n}\,r_{n_s}[A]\ldots r_{n_2}[A]\,r_{n_1}[A]\ A
\end{equation}
from a decomposition into homogeneous pieces
\begin{equation}
R_g[A] \= r_1[A]\ +\ g\,r_2[A]\ +\ g^2 r_3[A]\ +\ldots 
\qquad\textrm{with}\qquad \smallint A\tfrac{\delta}{\delta A}\,r_k[A] = k\,r_k[A]\ .
\end{equation}

Let us finally look a the diagrammatics in the Landau gauge~\cite{FL}.
With the solid line representing the free fermion propagator $(\im\slashed\pa)^{-1}$
and the dashed line standing for the free ghost propagator $\Box^{-1}$, we obtain the tree expansion \\[2pt]
\centerline{\includegraphics[width=0.7\paperwidth]{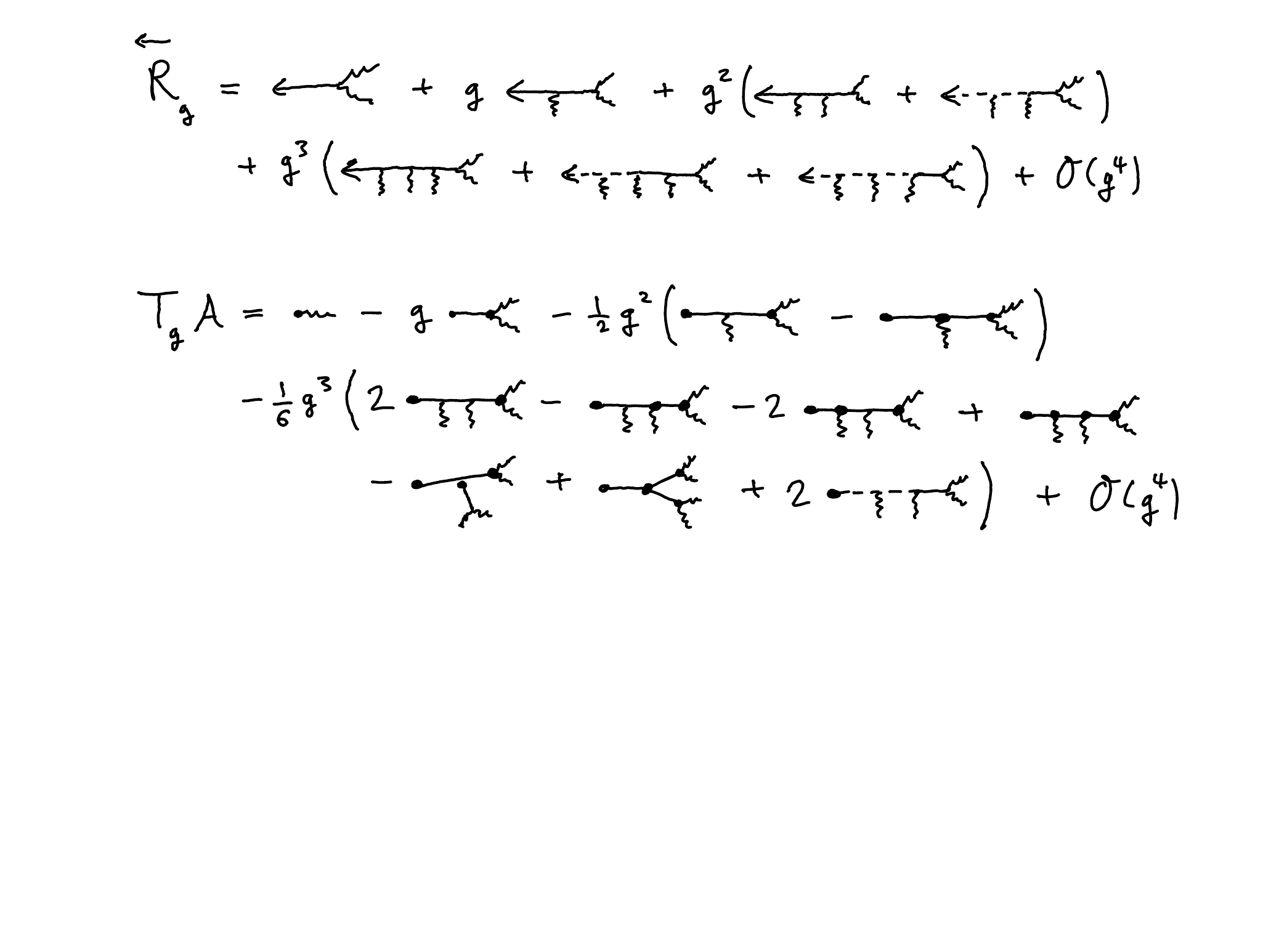}} \\
Iterating this in the universal formula~(\ref{universalA}) produces 
(with rules analogous to the scalar case) \\[4pt]
\centerline{\includegraphics[width=0.7\paperwidth]{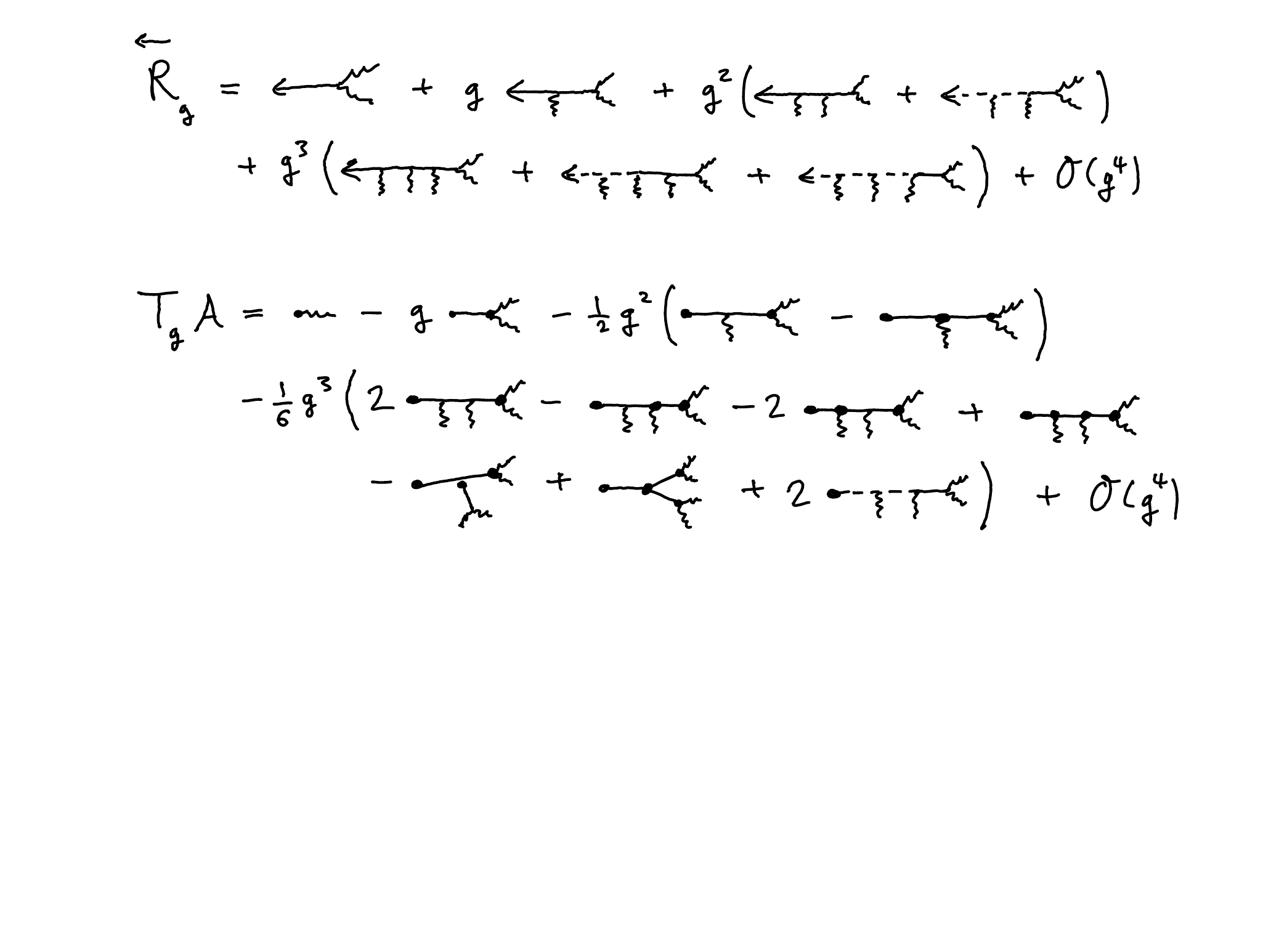}} \\
where the color structure follows the graphical one, 
and we have suppressed the Lorentz and spinor indices.
In fact, performing the spin traces creates various contractions
of Lorentz indices on the gauge-field legs and on the propagators due to
$(\im\slashed{\pa})^{-1}\!=\im\gamma^\mu\pa_\mu\Box^{-1}$, 
so that the number of terms at $O(g^n)$ grows rapidly with~$n$,
namely $1, 3, 34, 344, \ldots$.
Nevertheless, the expansion is algorithmic and may be implemented on a computer.
Explicit computations were performed to order~$g^3$ in~\cite{ALMNPP} 
and to order~$g^4$ in~\cite{MN}.
For first evaluations of correlators, see~\cite{DL2,NP}.

\section{Application to the supermembrane}
\noindent
In the last part of this talk I would like to describe a recent application~\cite{LN} of the Nicolai map
towards a quantization of the maximal supersymmetric membrane, 
an outstanding unsolved problem.\footnote{
See also H.~Nicolai's talk at the 
Humboldt Kolleg on Quantum Gravity and Fundamental Interactions,
which was part of the same Corfu Summer Institute 2021.}
The $D{=}11$ supermembrane~\cite{BST} can be obtained as an $N{\to}\infty$ limit of a 
maximally supersymmetric (so-called BFSS) matrix model~\cite{CH,BRR,Flume,dWHN,BFSS}. 
More concretely,
in a Minkowski background in the light-cone gauge, the supermembrane can be viewed as
a one-dimensional gauge theory of area-preserving diffeomorphisms (APD),
which is regularized by the SU($N$) BFSS matrix model. This matrix model arises also
in two other ways. Firstly, it can be seen as the worldline theory of a large number of $D0$-branes
in type IIA~string theory (the double dimensional reduction of the supermembrane).
Secondly, it appears as the Kaluza--Klein compactification of super Yang--Mills theory
from $1{+}9$ to $1{+}0$ dimensions. The Yang--Mills, matrix-model and APD coupling~$g$
can be seen to be proportional to the membrane tension~$T$, which combines the two
key parameters of string theory via $T=g_s^{-2/3}(\alpha')^{-1}$. Hence, a perturbative
quantization of the BFSS matrix model (in powers of~$g$) can serve as a low-$T$ expansion
of the quantum supermembrane. Here, we attempt to set this up via the Nicolai map,
by dimensionally reducing its $D{=}10$ SU($N$) super Yang--Mills version to a map for 
the matrix quantum mechanics and finally (in the $N{\to}\infty$ limit) for the APD gauge theory.

The Nicolai map for super Yang--Mills theory was described in the previous section.
Let us allow for $D=3,4,6$ or~$10$. 
The dimensional reduction from $\RR^{1,D-1}$ to $\RR^{1,0}$ effects
\begin{equation}
\pa_\mu \to (\pa_t, 0)\ ,\quad
A_\mu^A \to (\omega^A, X_a^A)\ ,\quad
\lambda_\alpha^A \to \theta_\alpha^A\ ,\quad
D_\mu \to (D_t{=}\pa_t{+}g\omega{\times}\,,\ g X_a^A{\times})
\end{equation}
where $\mu=(0,a)=(0,1,\ldots,D{-}1)$, $\alpha=1,\ldots,r$ and $A=1,\ldots,N^2{-}1$.
We use the $\times$ symbol to hide the SU($N$) structure constants, 
as in $(\omega{\times})^{AB}\equiv f^{ACB}\omega^C$.
The spinor index notation is a bit sloppy here: 
while $\lambda^A$ is an SO($D$) Majorana spinor, 
the SO($D{-}1$) Majorana $\theta^A$ has only half as many components
(the other half gets projected out).
The non-dynamical Lagrange multiplier~$\omega^A$ enforces the Gau\ss\ constraint.
The Lorenz gauge simplifies to
\begin{equation}
\cG(A)=\pa{\cdot}A \quad\longrightarrow\quad
\cG(\omega) = \dot\omega \equiv \pa_t\omega = D_t\omega\ ,
\end{equation}
thus $\cG{=}0$ forces $\omega$ to be constant in time.
Interestingly, the reduced temporal gauge $\omega{=}0$ implies 
the reduced Lorenz gauge $\dot\omega{=}0$.
Hiding color, Lorentz and spin indices, and integrating out the auxiliary $D$~field,
the Yang--Mills lagrangian reduces as follows,
\begin{equation}
\begin{aligned}
{\cal L}_{\textrm{YM}} &\= -\tfrac14 F^2-\tfrac{1}{2\xi}\cG(A)^2
-\tfrac{\im}{2}\bar\lambda\slashed{D}\lambda + \bar{c}\,\tfrac{\pa\cG}{\pa A}Dc
\quad\longrightarrow\\
{\cal L}_{\textrm{MM}} &\= \tfrac12(D_t X)^2 -\tfrac14 g^2(X{\times}X)^2
- \tfrac{\im}{2}\theta\,(D_t+g\gamma{\cdot}X{\times})\,\theta
- \tfrac{1}{2\xi}\dot\omega^2 + \bar{c}\,\pa_tD_t c\ .
\end{aligned}
\end{equation}

To construct the coupling flow operator for the matrix model, we may either
employ version~A of the previous section or directly dimensionally reduce
the Yang--Mills flow operator already given there. Either way, one arrives at
\begin{equation}
\overleftarrow{R}_g \= -\tfrac{1}{r}\smallint\!\!\smallint\!\!\smallint
\overleftarrow{\tfrac{\delta}{\delta X_a}} \Bigl[
\bigl(\underbrace{\gamma_a\unity - g X_a{\times}D_t^{-1}}_{\textrm{from}\ \ P_\mu^{\ \nu}}\bigr)\,
\bcontraction{}{\theta}{\ }{\theta}\theta\ \theta\,
\bigl(\underbrace{\tfrac12\gamma^{cd} X_c{\times}X_d + \gamma^d \omega{\times}
X_d}_{\textrm{from}\ \ \gamma^{\rho\sigma}A_\rho A_\sigma}\bigr) \Bigr]_{\alpha\alpha}
\end{equation}
where the Euclidean indices $a,b,\ldots=1,\ldots,D{-}1$ and the spin trace $[\ldots]_{\alpha\alpha}$
have been exhibited but color and the temporal argument in $X^A_a(t)$ are suppressed.
This operator is to be iterated on~$X$ to yield~$(T_gX)^A_a(t)$. 
Since no $\tfrac{\delta}{\delta\omega}$ appears, $R_g\omega{=}0$,
and hence $T_g\omega=\omega$ respects the gauge slice.
For simplicity, we pass to the temporal subgauge~$\omega{\equiv}0$.
Then, only odd powers of~$g$ show up in the perturbative expansion of~$R_g$.\footnote{
At least up to eighth order, where a nonzero contribution 
$\sim(\gamma_{a_1}\cdots\gamma_{a_9})_{\alpha\alpha}\sim\varepsilon_{a_1\cdots a_9}$ 
seems possible.}

With a solid line now depicting the one-dimensional propagator
$\pa_t^{-1}=\tfrac12\textrm{sgn}(t)=:\epsilon(t)$ up to a constant 
(and a linear term in case of a zero mode on a circle), 
the diagrammatical expansion of the flow operator reads\\
\centerline{\includegraphics[width=0.7\paperwidth]{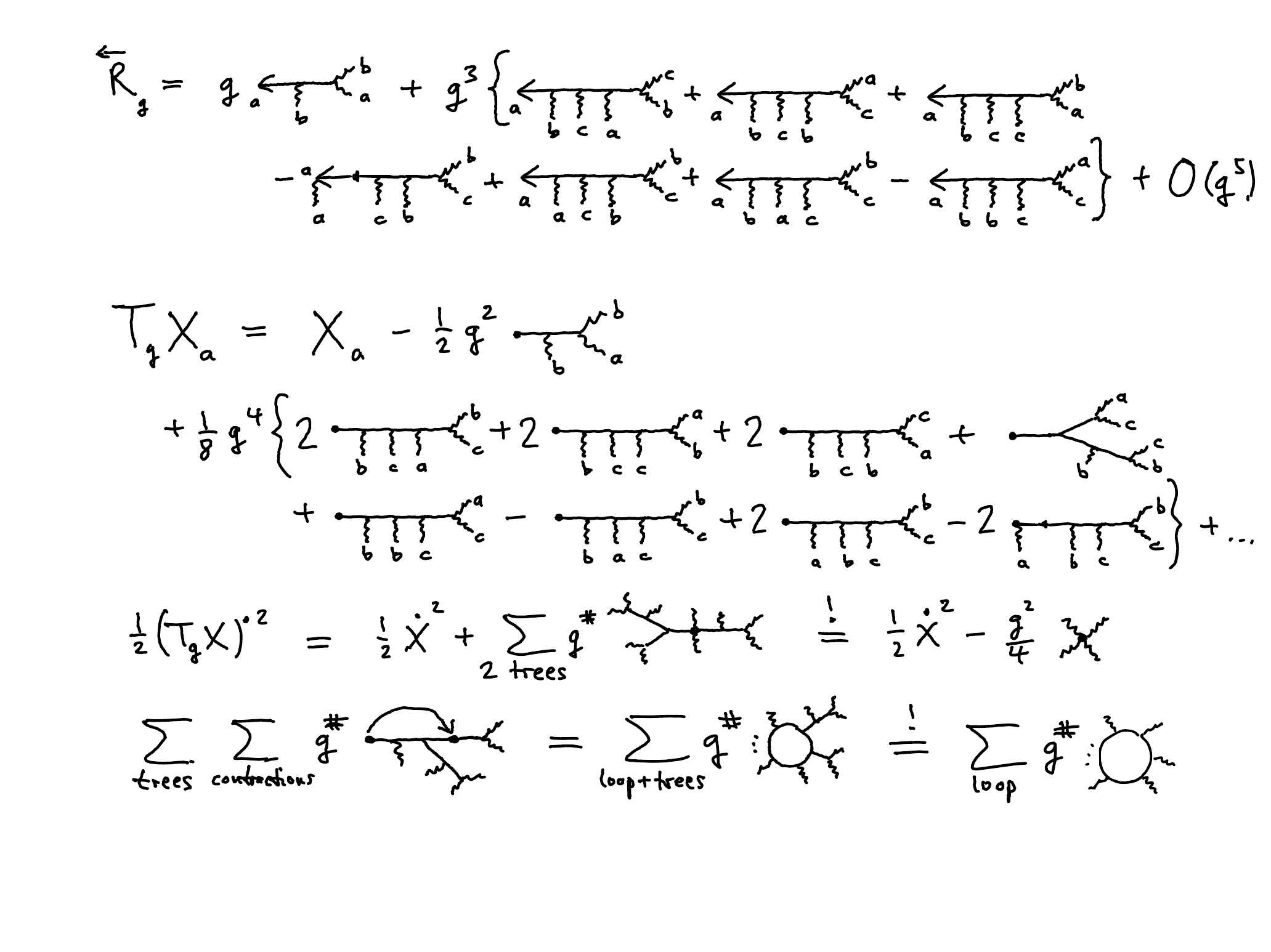}} \\
giving rise to the branched-tree expansion\\[4pt]
\centerline{\includegraphics[width=0.7\paperwidth]{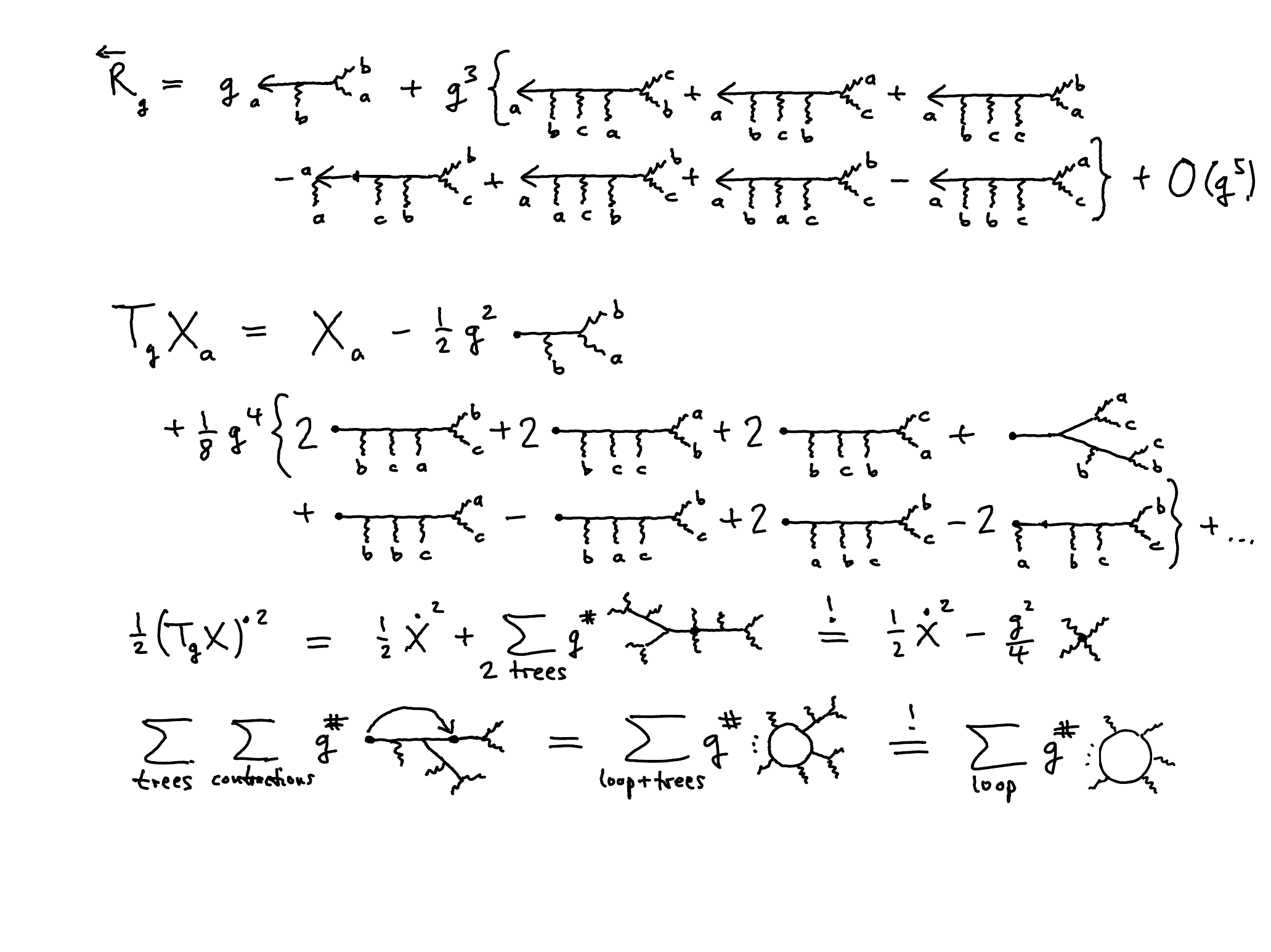}} \\
Remarkably, this expression passes all tests.
The free-action condition is met for any value of~$D$,\\[4pt]
\centerline{\includegraphics[width=0.7\paperwidth]{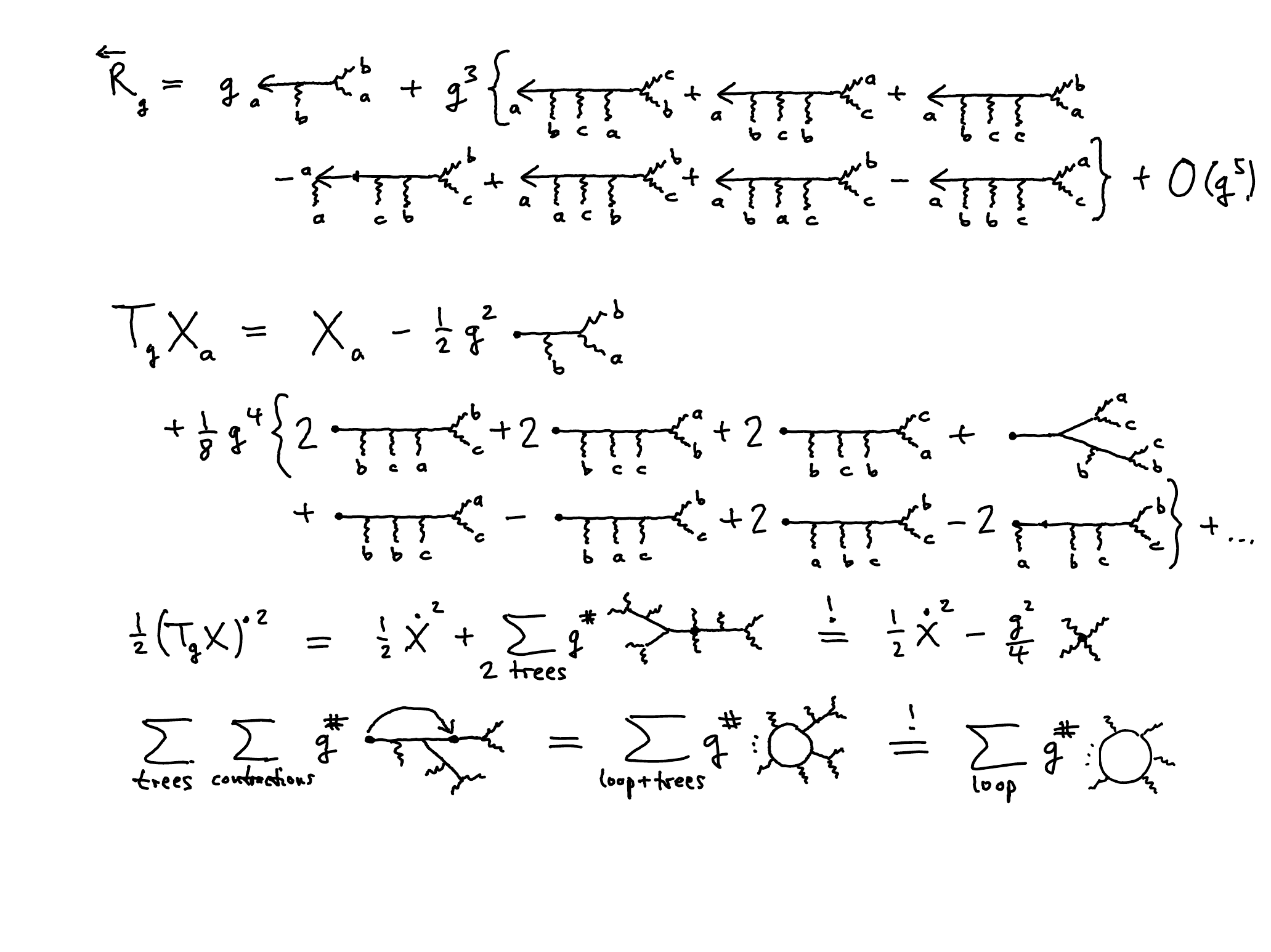}} \\
where $\#$ stands for the various $g$-powers in the sum.
It is nontrivial that all but one term cancel in the infinite sum over double trees.
The determinant matching, in contrast, works only for $D\in\{3,4,6,10\}$,\\
\centerline{\includegraphics[width=0.7\paperwidth]{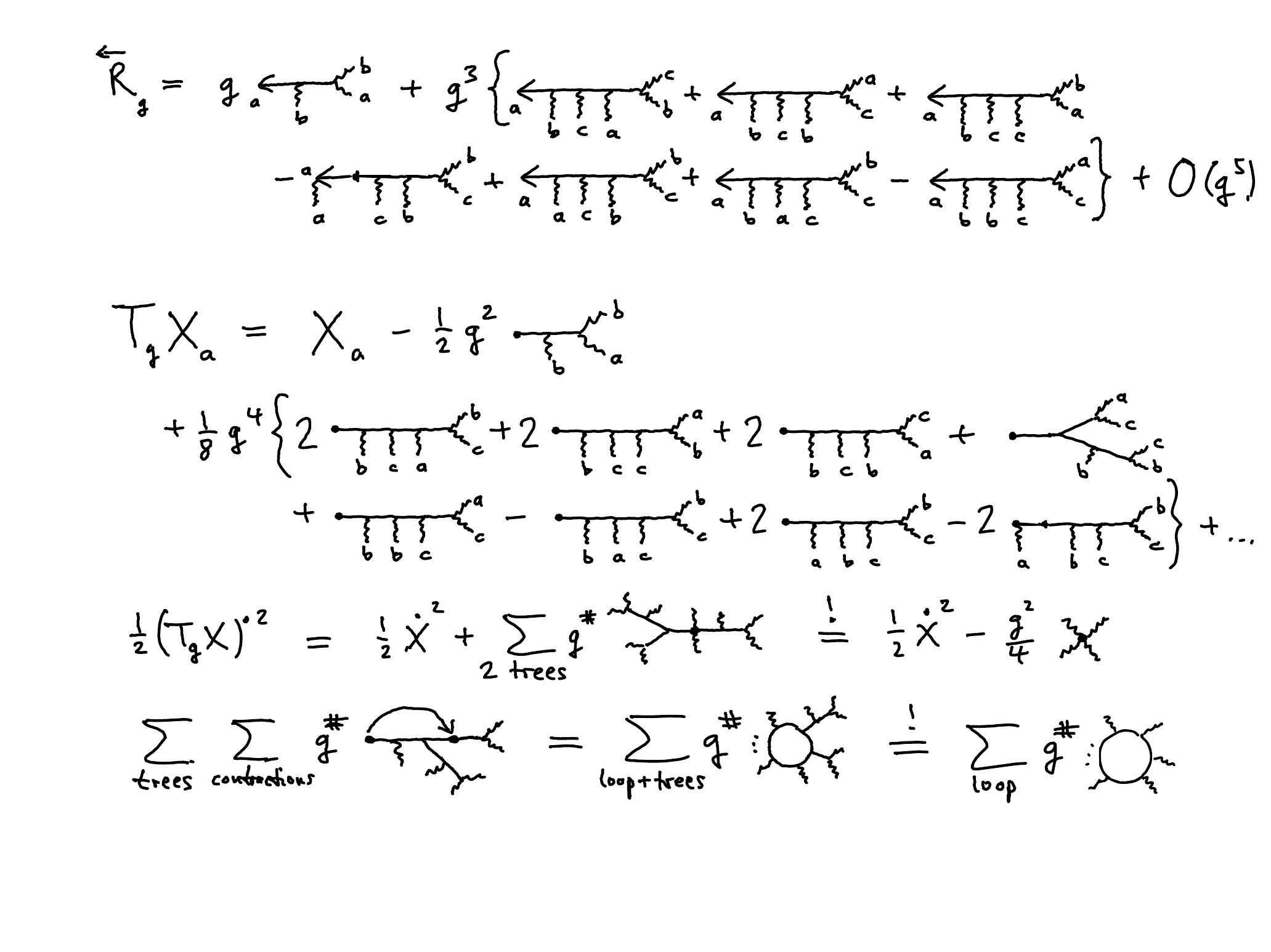}} \\
Here, it is amazing that (with the help of the Jacobi identity)
all loops with trees attached cancel out, leaving only the standard one-loop graphs.

The $N{\to}\infty$ limit leads to the area-preserving-diffeomorphism (APD) gauge theory,
\begin{equation}
X^A_a(t) \to X_a(\vec{\sigma},t) \quad\und\quad
f^{ABC} \to \smallint\diff^2\!\sigma\,\sqrt{w(\vec\sigma)}\
Y^A(\vec\sigma)\,\bigl\{Y^B(\vec\sigma)\,,Y^C(\vec\sigma)\bigr\}\ ,
\end{equation}
with membrane coordinates $\vec{\sigma}=(\sigma^1,\sigma^2)$,
a complete orthonormal basis $\bigl\{Y^A(\vec\sigma)\bigr\}$ of functions on the membrane,
and an irrelevant reference density~$w(\vec\sigma)$, which cancels when inserting the APD bracket
\begin{equation}
\bigl\{A(\vec\sigma)\,,B(\vec\sigma)\bigr\} \= \tfrac{1}{\sqrt{w(\vec\sigma)}}\,
\bigl( \pa_{\sigma^1} A(\vec\sigma)\,\pa_{\sigma^2} B(\vec\sigma)
- \pa_{\sigma^2} A(\vec\sigma)\,\pa_{\sigma^1} B(\vec\sigma) \bigr)\ .
\end{equation}
Using the $Y$~basis, the $N{\to}\infty$ limit of the color summation is converted into 
an integral over~$\vec\sigma$, and the matrix interaction gets encoded in the APD bracket, e.g.
\begin{equation}
f^{ABC} X^B_b(t) X^C_c(t) \quad\longrightarrow\quad
\bigl\{ X_b(\vec{\sigma},t)\,, X_c(\vec{\sigma},t) \bigr\}\ .
\end{equation}
This limit carries some subtleties. In particular, the APD bracket produces derivative 
(in $\sigma$) interactions, which may require a point-splitting regularization. 
In contrast, the absence of $\sigma$~derivatives in the quadratic part of the APD action
renders the latter ultralocal. This leads to singular $\delta(\vec{\sigma}{-}\vec{\sigma})$
factors in the fermion determinant which, however, cancel against like factors
in the Jacobian of the Nicolai map. Therefore, the the map remains well-defined 
in the large-$N$ limit because supersymmetry reigns!\footnote{
Conversely, it explains why this limit does not exist for the purely bosonic matrix model,
and why the bosonic membrane is `non-renormalizable'.}
More annoyingly, when the Nicolai map is employed in the perturbative computation of
APD correlators (e.g.~for membrane vertex operators), the ultralocal free propagator 
will lead to singularities $\sim\delta(\vec{0})^{-1}$.\footnote{
We thank J.~Plefka for this remark.}
This suggests that a partial resummation
is needed to pass from a worldline propagator to a membrane world-volume propagator,
in analogy with a geometric sum over mass insertions to shift from a massless propagator
to a massive one.
In APD language and suppressing the common $\vec\sigma$ arguments, 
the $N{\to}\infty$ limit of the above Nicolai map takes the form
\begin{equation}
\begin{aligned}
T_g X_a (t) \=\ X_a(t) &\ -\  \tfrac12 g^2 \int\!\!\!\!\int\!\!\md s\,\md u\; \ve(t{-}s)\,\ve(s{-}u)\,
\Big\{ X_b(s) \,,\,\bigl\{ X_b(u) , X_a(u) \bigl\} \Big\} \\
&\ +\ \tfrac18 g^4 \int\!\!\!\!\int\!\!\!\!\int\!\!\!\!\int\!\!\md s\,\md u\,\md v\,\md w\; 
\ve(t{-}s)\,\ve(s{-}u)\,\ve(u{-}v)\,\ve(v{-}w) \\
&\qquad\times\ \Biggl[ \ 
6\ \biggl\{ X_b(s)\,,\,\Bigl\{ X_c(u)\,,\,\bigl\{ X_{[a}(v)\,,\,\{ X_b(w),X_{c]}(w)\}\bigr\}\Bigr\}\biggr\} \\
&\qquad\ \  +\ 2\ \biggl\{ X_b(s)\,,\,\Bigl\{ X_{[b}(u)\,,\,\bigl\{ X_{|c|}(v)\,,\,\{ X_{a]}(w),X_c(w)\}\bigr\}\Bigr\}\biggr\} \\
&\qquad\ \  +\ 2\ \biggl\{ X_a(s)-X_a(t)\,,\,\Bigl\{ X_b(u)\,,\,\bigl\{ X_c(v)\,,\,\{ X_b(w),X_c(w)\}\bigr\}\Bigr\}\biggr\}\,
\Biggr] \\
&\quad +\ \tfrac18 g^4 \int\!\!\!\!\int\!\!\!\!\int\!\!\!\!\int\!\!\md s\,\md u\,\md v\,\md w\; 
\ve(t{-}s)\,\ve(s{-}u)\,\ve(s{-}v)\,\ve(v{-}w) \\
&\qquad\quad\times\ \biggl\{\bigl\{X_a(u),X_b(u)\bigr\}\,,\,\Bigl\{X_c(v)\,,\,\bigl\{X_b(w),X_c(w)\bigr\}\Bigr\}\biggr\}
\ \ +\ O(g^6)\ .
\end{aligned}
\end{equation}
By computer, this expression can easily be continued to any desired order in the coupling.

\section{Outlook}
\noindent
We have proposed a new angle of attack on the supermembrane, based on the Nicolai map
for the APD gauge theory. The perturbative small-tension expansion offers a path to quantization.
A distant goal is to establish quantum target-space Lorentz invariance for the supermembrane.
Closer in reach appears a computation of physically relevant correlation functions, 
e.g.~of graviton-emission vertex operators~\cite{DNP}
\begin{equation}
\begin{aligned}
V_h[X,\theta;k] \,=\, h^{ab} \Bigl[ 
& D_t X_a\,D_t X_b - \{X_a,\!X_c\}\{X_b,\!X^c\} - \im\bar\theta\gamma_a\{X_b,\!\theta\} \\
- \tfrac12 & D_t X_a\bar\theta\gamma_{bc}\theta k^c
- \tfrac12\{X_a,\!X^c\}\bar\theta\gamma_{bcd}\theta\,k^d 
+ \tfrac12\bar\theta\gamma_{ac}\theta\,\bar\theta\gamma_{bd}\theta\,k^c\!k^d \Bigr]
\,\ep^{-\im \vec{k}{\cdot}\vec{X} +\im k^-t}
\end{aligned}
\end{equation}
with graviton polarization~$h_{ab}$.
Another perspective is a control over the convergence of the perturbation series 
with the help of the universal formula~(\ref{universal}) for the map.
Puzzling is the special r\^ole of the Landau gauge for spacetime dimensions beyond four.
Finally, it would be marvellous to detect traces of ``integrability'' for maximally supersymmetric
Yang--Mills theory in four dimensions.



\begin{thebibliography}{99}
\addtolength{\itemsep}{-3.5pt}

\bibitem{Nic1}
H.~Nicolai,
{\it On a new characterization of scalar supersymmetric theories},\\
\href{https://dx.doi.org/10.1016/0370-2693(80)90138-0}
{{\it Phys.\ Lett.\ B} {\bf 89} (1980) 341}.
		
\bibitem{Nic2}
H.~Nicolai,
{\it Supersymmetry and functional integration measures},\\
\href{https://dx.doi.org/10.1016/0550-3213(80)90460-5}
{{\it Nucl.\ Phys.\ B} {\bf176} (1980) 419}.

\bibitem{Nic3}
H.~Nicolai,
{\it Supersymmetric functional integration measures},\\
lectures delivered at the NATO Advanced Study Institute on Supersymmetry,\\
Bonn, Germany, 20--31 Aug 1984, 
\href{https://cds.cern.ch/record/155731?ln=en}
{pp.393--420, eds. K.~Dietz et. al., {\it Plenum Press} (1984)}.

\bibitem{L1}
O.~Lechtenfeld,
{\it Construction of the Nicolai mapping in supersymmetric field theories},\\
Ph.D.\ Thesis, Bonn University (1984),
\href{https://lib-extopc.kek.jp/preprints/PDF/2000/0030/0030157.pdf}
{internal report {\it BONN-IR-84-42}, ISSN-0172-8741}.
		
\bibitem{DL1}
K.~Dietz and O.~Lechtenfeld,
{\it Nicolai maps and stochastic observables from a coupling constant flow},
\href{https://dx.doi.org/10.1016/0550-3213(85)90132-4}
{{\it Nucl.\ Phys.\ B} {\bf 255} (1985) 149}.
		
\bibitem{DL2}
K.~Dietz and O.~Lechtenfeld,
{\it Ghost-free quantisation of non-Abelian gauge theories 
via the Nicolai transformation of their supersymmetric extensions},
\href{https://dx.doi.org/10.1016/0550-3213(85)90642-X}
{{\it Nucl.\ Phys.\ B} {\bf 259} (1985) 397}.

\bibitem{L2}
O.~Lechtenfeld,
{\it Stochastic variables in ten dimensions?},
\href{https://doi.org/10.1016/0550-3213(86)90531-6}
{{\it Nucl.\ Phys.\ B} {\bf 274} (1986) 633}.

\bibitem{LR1}
O.~Lechtenfeld and M.~Rupprecht,
{\it Universal form of the Nicolai map},\\
\href{https://doi.org/10.1103/PhysRevD.104.L021701}
{{\it Phys.\ Rev.\ D} {\bf 104} (2021) L021701}
[\href{https://arxiv.org/abs/2104.00012}{arXiv:2104.00012 [hep-th]}].

\bibitem{FL}
R.~Flume and O.~Lechtenfeld,
{\it On the stochastic structure of globally supersymmetric field theories},
\href{https://doi.org/10.1016/0370-2693(84)90459-3}
{{\it Phys.\ Lett.\ B} {\bf 135} (1984) 91}.

\bibitem{ALMNPP}
S.~Ananth, O.~Lechtenfeld, H.~Malcha, H.~Nicolai, C.~Pandey and S.~Pant,\\
{\it Perturbative linearization of supersymmetric Yang--Mills theory},\\
\href{https://dx.doi.org/10.1007/JHEP10(2020)199}
{{\it JHEP} {\bf 10} (2020) 199}
[\href{https://arxiv.org/abs/2005.12324}{arXiv:2005.12324 [hep-th]}].

\bibitem{BSS}
L. Brink, J.~H. Schwarz and J.~Scherk,
{\it Supersymmetric Yang--Mills theories},\\
\href{https://doi.org/10.1016/0550-3213(77)90328-5}
{{\it Nucl.\ Phys.\ B} {\bf121} (1977) 77}.

\bibitem{ANPP}
S.~Ananth, H.~Nicolai, C.~Pandey and S.~Pant,
{\it Supersymmetric Yang--Mills theories: not quite the usual perspective},
\href{https://doi.org/10.1088/1751-8121/ab7b9d}
{{\it J.\ Phys.\ A: Math.\ Theor.} {\bf 53} (2020) 174001}
[\href{https://arxiv.org/abs/2001.02768}{arXiv:2001.02768 [hep-th]}].

\bibitem{MN}
H.~Malcha and H.~Nicolai,
{\it Perturbative linearization of super-Yang--Mills theories in general gauges},
\href{https://doi.org/10.1007/JHEP06(2021)001}
{{\it JHEP} {\bf 06} (2021) 001}
[\href{https://arxiv.org/abs/2104.06017}{arXiv:2104.06017 [hep-th]}].

\bibitem{LR2}
O.~Lechtenfeld and M.~Rupprecht,
{\it Construction method for the Nicolai map in supersymmetric Yang--Mills theories},
\href{https://doi.org/10.1016/j.physletb.2021.136413}
{{\it Phys.\ Lett.\ B} {\bf 819} (2021) 136413}
[\href{https://arxiv.org/abs/2104.09654}{arXiv:2104.09654 [hep-th]}].

\bibitem{NP} 
H.~Nicolai and J.~Plefka, 
{\it $N{=}4$ super-Yang--Mills correlators without anticommuting variables},
\href{https://doi.org/10.1103/PhysRevD.101.125013}
{{\it Phys.\ Rev.\ D}  {\bf 101} (2020) 125013}
[\href{https://arxiv.org/abs/2003.14325}{arXiv:2003.14325 [hep-th]}].

\bibitem{LN}
O.~Lechtenfeld and H.~Nicolai,
{\it A perturbative expansion scheme for supermembrane and matrix theory},
\href{https://doi.org/10.1007/JHEP02(2022)114}
{{\it JHEP} {\bf 02} (2022) 114}
[\href{https://arxiv.org/abs/2109.00346}{arXiv:2109.00346 [hep-th]}].

\bibitem{BST}
E.~Bergshoeff, E.~Sezgin and P.K.~Townsend, \\
{\it Supermembranes and eleven-dimensional supergravity},
\href{https://doi.org/10.1016/0370-2693(87)91272-X}
{{\it Phys.\ Lett.\ B} {\bf 189} (1987) 75}; \\
{\it Properties of the eleven-dimensional supermembrane theory},
\href{https://doi.org/10.1016/0003-4916(88)90050-4}
{{\it Ann.\ Phys.} {\bf 185} (1988) 330}.

\bibitem{CH}
M.~Claudson and M.B.~Halpern, 
{\it Supersymmetric ground state wave functions},\\
\href{https://doi.org/10.1016/0550-3213(85)90500-0}
{{\it Nucl.\ Phys.\ B} {\bf 250} (1985) 689}.

\bibitem{BRR}
M.~Baake, M.~Reinicke and V.~Rittenberg, 
{\it Fierz identities for real Clifford algebras and the number of supercharges}, 
\href{https://doi.org/10.1063/1.526539}
{{\it Journ.\ Math.\ Phys.} {\bf 26} (1985) 1070}.

\bibitem{Flume}
R.~Flume, 
{\it On quantum mechanics with extended supersymmetry and nonabelian gauge constraints},
\href{https://doi.org/10.1016/0003-4916(85)90008-9}
{{\it Ann.\ Phys.} {\bf 164} (1985) 189}.

\bibitem{dWHN}
B.~de Wit, J.~Hoppe and H.~Nicolai,
{\it On the quantum mechanics of supermembranes}, \\
\href{https://doi.org/10.1016/0550-3213(88)90116-2}
{{\it Nucl.\ Phys.\ B} {\bf 305} (1988) 545.}

\bibitem{BFSS}
T.~Banks, W.~Fischler, S.H.~Shenker and L.~Susskind,
{\it M theory as a matrix model: a conjecture},\\
\href{https://doi.org/10.1103/PhysRevD.55.5112}
{{\it Phys.\ Rev.\ D} {\bf 55} (1997) 5112} 
[\href{https://arxiv.org/abs/hep-th/9610043}{arXiv:hep-th/9610043 [hep-th]}].

\bibitem{DNP} 
A.~Dasgupta, H.~Nicolai and J.C.~Plefka, 
{\it Vertex operators for the supermembrane},\\
\href{https://doi.org/10.1088/1126-6708/2000/05/007}
{{\it JHEP} {\bf 05} (2000) 007}
[\href{https://arxiv.org/abs/hep-th/0003280}{arXiv:hep-th/0003280 hep-th]}].
	
\end{thebibliography}
\end{document}